\documentclass[12pt,preprint]{aastex}
\shorttitle{A flux tube solar dynamo model}
\newcommand{\p}[2]{\frac{\partial #1}{\partial #2}}
\newcommand{\pp}[2]{\frac{\partial^2 #1}{\partial #2^2}}

\newcommand{\vb}[1]{{\bf #1}}
\newcommand{\s}[1]{\mbox{\scriptsize{#1}}}
\newcommand{\h}[1]{{\cal #1}}

\begin{document}

\title{A flux tube solar dynamo model based on the competing role of buoyancy and downflows}

\author{L. H. Li, S. Sofia}
\affil{Department of Astronomy, Yale University, P.O. Box 208101, New Haven, CT 06520-8101}
\email{li@astro.yale.edu}
\author{G. Belvedere}
\affil{Astrophysics Section, Department of Physics and Astronomy, University of Catania, 
Via Santa Sofia 78, 95123 Catania, Italy}

\begin{abstract}
A magnetic flux tube may be considered both as a separate body and as a confined
field. As a field, it is affected both by differential rotation ($\Omega$-effect)
and cyclonic convection ($\alpha$-effect). As a body, the tube experiences not
only a buoyant force, but also a dynamic pressure due to downflows above the
tube. These two competing dynamic effects are incorporated into the
$\alpha$-$\Omega$ dynamo equations through the total
magnetic turbulent diffusivity, leading to a flux tube dynamo operating in the
convection zone. 
We analyze and solve the extended dynamo equations in the linear approximation
by adopting the observed solar internal rotation and assuming a downflow effect
derived from numerical simulations of solar convection zone.
The model reproduces: the 22-year cycle period; the extended butterfly diagram
with the confinement of strong activity to low heliographic latitudes $|\Phi|\le
35^\circ$; the evidence that at low latitudes the radial field is in an
approximately $\pi$ phase lag compared to the toroidal field at the same
latitude; the evidence that the poleward branch is in a $\pi/2$ phase lag with
respect to the equatorward branch; and the evidence that most of the magnetic flux
is present in an intermittent form, concentrated into strong flux tubes. 
\end{abstract}


\maketitle

\section{Introduction}\label{sect:s0}

The magnetic activity of the Sun and late type stars is currently attributed to
a dynamo action, namely the interaction of conducting fluid motions and
magnetic fields in a suitable spherical shell. This was first suggested by
Larmor (1919), but the development of these ideas proved to be very complex, especially in view of
Cowling's theorem (1934), which indicated that a
dynamo could not operate with strictly axisymmetric magnetic fields. The generation and evolution
of magnetic fields in the Sun is mostly described in terms of an $\alpha$-$\Omega$
dynamo operating in the convection zone or in the boundary (overshoot) layer
just beneath the convection zone, as the well known recent results of
helioseismology may suggest. Although the physics of the interaction of rotation
and convective modes is very complex, and poorly understood in spite of ever more
refined attempts, the simple (in principle) approach based on differential
rotation and cyclonic turbulence has been productive in modeling solar
activity.

A good dynamo model should, of course, reproduce the main characteristics of the
solar cycle: equatorward migration of spots and, in general, of active regions
where strong magnetic fields are observed, starting from a latitude of about
35$^\circ$; poleward migration of some magnetic tracers such as high latitude
prominences and polar faculae (e.g., Makarov \&\ Sivaraman 1989); polarity 
reversals of the polar fields every 11-years; polarity rules in the two hemispheres 
(Hale's law); ratio and phase 
relation of the poloidal field to the toroidal field (\cite{S76}); 22-year cycle period;
intermittent form of most of the magnetic flux, which is mainly concentrated
into flux tubes; rigid rotation of the preferential longitudes (some fixed
longitudes where spots show a rigid rotation); flip-flop effect in the
preferential longitudes (namely the alternate occurrence of spots at these
longitudes in the two 11-year halves of the 22-year magnetic cycle); and possibly
long-term periodicities shown by solar activity, and absence of activity at some
epochs (Maunder minima).

The location of the $\alpha$-$\Omega$ dynamo action is an important theoretical
problem. In the course of the years, several locations have been proposed:
mostly the whole convection zone or the stable layers at the bottom of the
convection zone (this has been done in an extensive series of papers starting
from the fundamental work by Parker (1955)); but also the boundary (overshoot
layer) just beneath the convection zone (e.g., Belvedere et al. 1991,
R\"{u}diger \&\ Brandenburg 1995). It is usually assumed that the $\Omega$ and
$\alpha$-effects are located in the same region, since a spatial separation is
not plausible as would imply problems of upward and downward magnetic field
transport which are not easily overcome by means of the turbulent diffusion. An
exception are the so-called interface models \cite{P93,CM97,MT99}, that assume
the $\Omega$-effect to operate on the low diffusivity side of an interface (the
base of the convection zone), while the $\alpha$-effect works on the high
diffusivity side, and, in some aspects, also the advective models
\cite{CSD95,D95,D96,D97,NC01}, see also the recent work by Bonanno et al.
(2002). In these models, the location of the two effects may be varied,
sometimes having the characteristics of an interface dynamo, sometimes of a
Babcock-Leighton dynamo  (with a deeply located $\Omega$-effect and a surface
located $\alpha$-effect). The advection dynamos essentially work like a conveyor
belt that is provided by the meridian circulation in the form of a single or
more cells in the convection zone (Wang \&\ Sheeley 1991; Dikpati \&\ Charbonneau 1999).

Of course the validity of the various approaches and models is measured by their
capability to reproduce as many features of the solar activity as possible. But this
is not the only criterion to evaluate a dynamo model: in fact, the
current framework of solar dynamo theory, which appears to be in 
agreement with the basic features of solar activity, arose from a series of
attempts that dealt separately with partial, although fundamental aspects of the problem.

Among those attempts, in the framework of this
paper, we consider flux tube dynamos. It is likely that, in future work on MHD
of stellar interiors and dynamo theory, flux tube dynamics will be 
particularly important. Earlier studies on the dynamics of thin flux tubes,
intermittently erupting at the solar surface, were carried out by (e.g.)
Schuessler (1980), Spruit (1981), Spruit and Van Ballegooijen (1982a,b), van
Ballegooijen (1983), Schmitt (1987), Ferriz-Mas and Schuessler(1994,1995),
Caligari et al.(1995), and Schuessler et al. (1996).

In the thin tube approximation, the tube radius must be much smaller than its
radius of curvature, the local scale height, the rotational shear scale, and the
size of the relevant perturbations, including the large scale turbulence size
(i.e. the size of dominant eddies). The basic point is that strong axisymmetric
toroidal fields located in the subadiabatic and relatively stable layers at the
bottom of the convection zone, generated from poloidal fields by rotational
shear, can be represented as an ensemble of isolated flux tubes, subjected to
magnetic buoyancy and other perturbations like aerodynamical drag, downward
flows and rotationally induced effects such as Coriolis force and differential
centrifugal force.
 
In this paper, we argue that, due to the competition between
two effects, namely buoyancy force and downward motions, the rising time of the
tubes may be comparable to the timescale of the solar cycle. In fact, numerical
simulations of the outer solar convection zone indicate a major presence of
downward moving plumes \citep{CS89,KC98,N99,RDLSKCG03}.
A similar idea was already presented by Parker (1975), who considered the
aerodynamical drag as the competing effect.
The new approach we adopt in this paper consists in giving a particular
relevance to the downward flows and in incorporating the two competing effects
into the mean field dynamo equations, by deriving an expression for the total
turbulent diffusivity that takes into account the action of both buoyancy and
downward motions. Then, by solving the linear problem in the usual perturbative 
methods, we obtain, in a
self-consisting way, results in close agreement with the main observed solar
cycle features. A
good feature of this model is the self-consistency of the approach formulated in the
framework of a single
mathematical (and physical) development, without any ad hoc parameterizations.

In this model the solar internal differential rotation deduced from inversions
of helioseismic observations (e.g., Antia et al. 1998) has been adopted.

We show in \S\ref{sect:s1} how the effects of magnetic buoyancy and downward
motions can be incorporated into the mean field dynamo equations by modifying
the usual expression for the turbulent diffusivity $\beta$. In \S\ref{sect:s2}
we linearly analyze the dynamo equations to obtain the observable quantities of
the model. In \S\ref{sect:s3} we use the theoretical findings to interpret 
the properties of the solar cycle listed above. Finally, in \S\ref{sect:s4} 
we describe the main conclusions of the paper.

Notice that in this analysis antisymmetrical (dipole-like) modes are naturally
assumed, as observations show that they are dominant in the Sun. Indeed, it is
beyond the scope of this paper to solve a bidimensional ($r,\theta$) eigenvalue
problem in order to check whether dipole-modes are more easily excited and why
(as was done, e.g., by Bonanno et al. 2002).

\section{Basic equations}\label{sect:s1}

We consider a magnetic flux tube with radius $a$ and length $L$. Its magnetic field 
$\vb{B}$ is assumed to be uniform. As usual, we decompose it into a toroidal and a 
poloidal component,
\begin{equation}
  \vb{B} = B \vb{e}_{\phi} + \nabla\times (A\vb{e}_{\phi}), \label{eq:b}
\end{equation}
where $\vb{e}_\phi$ is the azimuthal unit vector, $B$ is the toroidal field component, 
and $A$ is the (toroidal) vector potential of the poloidal field. The local hydrostatic 
equilibrium of the flux tube in the solar interior requires
\begin{equation}
P = P_i+B^2/8\pi,
\end{equation}
where $P$ and $P_i$ are the total and internal gas pressure, respectively. Assuming 
that this equilibrium in the flux tube is reached solely by adjusting its density, 
then the internal density $\rho_i$ is related to the normal density $\rho$ by
\begin{equation}
  \rho_i = \rho P_i/P.
\end{equation}
As a result, the density reduction of the magnetic flux tube is
\begin{equation}
 \rho-\rho_i=\rho B^2/8\pi P.
\end{equation}
The reduced density inside the flux tube produces a buoyancy,
\begin{equation}
  \vb{F} = \pi a^2L g(\rho-\rho_i) \vb{e}_r,
\end{equation}
where $g$ is the gravitational acceleration. We assume that $a$ is smaller than 
the pressure scale height so that we can consider that the gas density of the 
flux tube is uniform. In this case, the total mass contained in the flux tube equals
\begin{equation}
  M_i = \pi a^2L\rho(1-B^2/B_g),
\end{equation}
where $B_g=(8\pi P)^{1/2}$. 
Therefore, the buoyant acceleration $g_b$ can be expressed by the gravitational 
acceleration $g$, total pressure $P$, and the magnetic field strength $B$, as follows:
\begin{equation}
  g_b = gB^2/(B_g^2-B^2). \label{eq:gb}
\end{equation}

If there is no downward flow to balance this pressure, the only force that goes 
against the buoyant force is the aerodynamic drag,
\begin{equation}
  \vb{F}_d = -\case{1}{2}C_d\rho u^2 aL \vb{e}_r,
\end{equation}
where $u$ is the velocity of rise and $C_d (\sim 1)$ is the drag coefficient. 
Taking into account both the buoyant force and the drag force, as done by 
Parker (1975), the terminal velocity of rise occurs for $F=F_D$, yielding 
the rate of rise
\begin{equation}
   u = \left(\frac{2\pi a g}{C_d}\frac{\rho-\rho_i}{\rho}
  \right)^{1/2}=V_A\left(\frac{\pi a}{C_D H_p}\right)^{1/2},
\end{equation}
where $H_p=P/\rho g$ is the pressure scale height, $V_A=B/(4\pi\rho)^{1/2}$ 
is the Alfv\'{e}n speed. The rise time to the surface is smaller than the 
required field amplification time of about 5 years by the classical dynamo 
theory, as estimated by Parker (1975).

However, numerical simulations of the Sun's outer convection zone made, 
e.g. by Chan \&\ Sofia (1989), indicate a major presence of downward-moving 
plumes of high velocity. This downward flow has an inward dynamic pressure 
$\rho v_z^2$. If this pressure can balance the buoyant force of the flux 
tube, then the flux tube can be stored in the convection zone so that it 
can be amplified by the $\alpha$-$\Omega$ dynamo. As a body, 
a magnetic flux tube experiences not only a buoyancy, but also a dynamic 
pressure of the downflows. The total pushdown force of downward plumes 
equals to the dynamic pressure ($\rho v_z^2$) times the total cross 
section of the plumes ($S$). The resulting acceleration $g_v$ can be expressed as follows:
\begin{equation}
  g_v = 2s v_z^2 \xi/[\pi a(1-B^2/B_g^2)], \label{eq:gn}
\end{equation}
where $s=1$ when $v_z>0$, $s=-1$ when $v_z<0$, $\xi=S/2aL\le 1$ is the 
fractional area of the downflows. The upward-moving plumes underneath 
the flux tube ($s=1$) accelerate the magnetic buoyant diffusion of the 
tube, while the downward-moving plumes above the flux tube ($s=-1$) go 
against the magnetic diffusion. These downflows act as an anti-diffusion 
process. In general, $-1\le s \le +1$, depending on the localtion of the flux tube.

\begin{figure}
\plotone{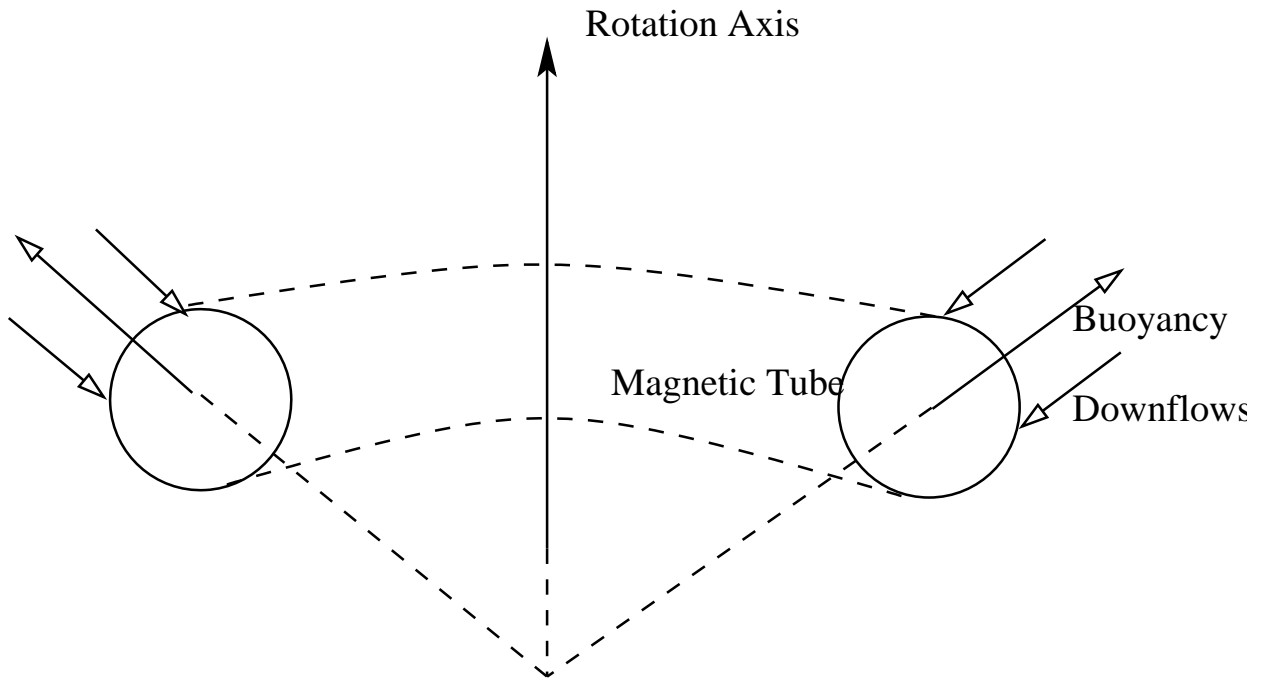}
\caption{
 Schematic of the force balance of a magnetic tube in the convection zone. 
\label{fig:f1}}
\end{figure}

The condition that the dynamic pressure of the downflows balances the buoyancy
determines a critical value of the magnetic field $B_c$, which is given by $g_b+g_v=0$,
\begin{equation}
  B_c = B_g (-2s v_z^2 \xi/\pi ag)^{1/2}. \label{eq:bc}
\end{equation}
When $B< B_c$, the buoyancy is smaller than the dynamic pressure of the downflows, 
the dynamo magnetic field may grow; when $B>B_c$, the dynamo field may damp. These 
two ingredients, buoyancy and downflows, are shown in Fig.~\ref{fig:f1}. Since downflows 
occur in the convection zone and the overshoot region near the base of the convection 
zone, this dynamo mechanism may operate in both of these regions.

The magnetic field in a flux tube can be considered to be a ``mean field'' 
since the mean-field concept is relative (e.g., 
Krause and  R\"{a}dler 1980). In fact, it depends on the 
spatio-temporal range over which the average is made. For the mean field 
in a flux tube, the spatio-temporal range is the volume and lifetime of 
a typical magnetic flux tube. With this in mind, the field in a flux tube
obeys the classic dynamo equation \cite{P55},
\begin{equation}
  \p{\vb{B}}{t} = \nabla\times(\vb{U}\times\vb{B} + \vec{{\cal E}}) -\nabla\times( \eta\nabla\times\vb{B}). \label{eq:dynamo}
\end{equation}
Using this equation, we cannot study the formation of flux tubes from the 
large-scale magnetic field, but we can determine if the flux tube fields can be
amplified. We assume a pure rotation motion $\vb{U}=\vb{\Omega}\times\vb{r}$, where $\vb{\Omega}$ is the angular velocity.
This will cause the $\Omega$-effect. 

Both the buoyant and dynamic diffusivities are direction-dependent. So they are anisotropic. Such anisotropic diffusivities must be treated as tensors in principle (Krause \&\ R\"{a}dler 1980). However, it is hard to do so in practice. To our knowledge, almost all dynamo models (including most conveyor belt ones) consider thermal, 
dynamical and magnetic diffusivities as scalars. All recent books that deal also with solar dynamo models (e.g., Stix 1989; Mestel 1999; Ruediger \&\ Hollerbach 2004) quote and review only papers that assume scalar diffusivities.
So, a solar model that includes the anisotropy of diffusivity in tensorial form is currently beyond the state of the art.

Physically, the tensorial form of diffusivity should be used only when a strong source of asymmetriy exists. This is not the stellar case, where spherical symmetry is generally assumed, and no strong axial electromotive force, no fast rotation occur. Practically, the dynamo theory that uses the scalar form of diffusivities has captured the essentials of the solar cycle, and is able to reproduce the butterfly diagram, and other observational facts. Mathematically, the scalar form of diffusivities can be considered to be the zero-order approximation of the general tensorial form. For these reasons, we use the following approximation form of the general electromotive force (Krause \&\ R\"{a}dler 1980)
\begin{equation}
  \h{E} \approx -\alpha^{(0)} \vb{B} - \beta^{(0)}\nabla\times\vb{B},
\end{equation}
where
\begin{equation}
  \beta^{(0)} \approx \eta + \beta_t+\beta_b+\beta_v.
\end{equation}
Defining
\begin{equation}
  \alpha\equiv-\alpha^{(0)}, \hspace{5mm} \beta\equiv \beta^{(0)},
\end{equation}
we obtain the governing equation of a magnetic flux tube,
\begin{equation}
  \p{\vb{B}}{t} = \nabla\times(\vb{U}\times\vb{B}+\alpha\vb{B}) -\nabla\times( \beta\nabla\times\vb{B}). \label{eq:model}
\end{equation}
Since $\beta$ depends on $\vb{B}$, this equation is nonlinear. In the azimuthal symmetric case, this equation is equivalent to the following two scalar equations:
\begin{eqnarray}
 \p{A}{t} &=& \alpha B + \beta\left( \frac{1}{r}\p{A}{r} + \frac{\cot\theta}{r^2}\p{A}{\theta} + \pp{A}{r} + \frac{1}{r^2}\pp{A}{\theta}\right), \nonumber\\ \label{eq:dynamo1} \\
 \p{B}{t} &=& a_1 A + a_2 \p{A}{r} + a_3 \p{A}{\theta} -\alpha \left(\pp{A}{r}+\frac{1}{r^2} \pp{A}{\theta}\right) \nonumber\\
 && + b_1 B + b_2 \p{B}{r} + b_3\p{B}{\theta}+ \beta \left(\pp{B}{r} + \frac{1}{r^2} \pp{B}{\theta}\right).\nonumber\\ \label{eq:dynamo2} 
\end{eqnarray}
See appendix~\ref{app:a} for the detailed derivation and the definitions for those coefficients $a_s$ and $b_s$. Here $B$ and $A$ are defined in Eq.~(\ref{eq:b}).

The magnetic diffusion caused by the microscopic gas collision motion, has been included in Eq.~(\ref{eq:dynamo}) in terms of the classic magnetic diffusivity $\eta\nabla\times\vb{B}$. For a homogeneous convection, $\alpha$ can be approximately expressed as \citep{KR80}
\begin{equation}
  \alpha = \case{1}{3} h \tau_{\s{cor}},
\end{equation}
where $h=\overline{\vb{v}\cdot(\nabla\times\vb{v})}$ is the mean helicity over the correlation time $\tau_{\s{cor}}$.
But in the general case, $\alpha$ may deviate from this expression. Since the turbulent velocity $\vb{v}$ is related to the background and dynamo magnetic field $\vb{B}$, $\alpha$ depends upon $\vb{B}$. This means that $\alpha$-effects must be nonlinear.

In order to work out the various magnetic diffusivities mentioned above, we observe that the diffusivity of a force is proportional to its impulse per unit mass. The impulse is defined as the force times its action time. The impulse per unit mass has the dimension of acceleration times time. The action time can be taken to be the dynamic response time of a magnetic tube. Since it is assumed that the process is iso-thermal in calculating the buoyant acceleration, the dynamic response time is the travel time of a perturbation across the 
diameter of the flux tube at the isothermal sound speed, $C_s=(P/\rho)^{1/2}$, so $t_{\s{act}}=2a/C_s$. However, diffusivity has the dimension of acceleration times time times length. The shortest characteristic length for a flux tube is its diameter $a$. We use this length $a$ to meet the need of dimensionality for the buoyant and the dynamic diffusivity. Consequently, we have
\begin{eqnarray}
 \beta_b &=& g_b (2a/C_s)2a=\frac{(\rho/P)^{1/2} 4a^2g  B^2}{B_g^2-B^2} , \label{eq:betab} \\
 \beta_v &=& g_v (2a/C_s)2a=\frac{(\rho/P)^{1/2}8asv_z^2\xi B_g^2}{\pi (B_g^2-B^2)}. \label{eq:betad}
\end{eqnarray}
The magnetic turbulent diffusivity $\beta_t=\case{1}{3} v'^2 \tau_{\s{cor}}$ for an isotropic turbulence is known \citep{KR80} where $v'$ is the turbulent velocity.

\section{Linear analysis}\label{sect:s2}

The flux tube model equations obtained in the previous section are nonlinear partial-differential equations. Since such equations are difficult to solve analytically, while a simple linear analysis may reveal some important characteristics about their solution, we analyze and solve the equation in the linear approximation in this paper. Unlike numerical simulations, the linear analysis cannot take into account full nonlinearity though it reveals some nonlinear effects such as critical magnetic fields of a flux tube.

For the sake of simplicity, we neglect $\eta$ in this analysis, namely, we assume
\begin{equation}
 \beta \approx \beta_t+\beta_b+\beta_v.
\end{equation}
The second simplification assumption is $B^2\ll B_g$, which means
\begin{eqnarray}
 \beta_b &\approx& (\rho/P)^{1/2} 4a^2g  B^2/B_g^2 , \\
 \beta_v &\approx& (\rho/P)^{1/2}8asv_z^2\xi/\pi.
\end{eqnarray}

\subsection{Linearized dynamo equations}

Since $b_s$ and $\beta$ in Eqs.~(\ref{eq:dynamo1}-\ref{eq:dynamo2}) depend upon $B$, $\partial B/\partial r$ and/or $\partial B/\partial\theta$, we need to specify $B=B_0$ and $A=A_0$ as the reference state to investigate their perturbations $B'$ and $A'$ near the reference state. See Appendix~\ref{app:b} for the detailed linearization of Eqs.~(\ref{eq:dynamo1}-\ref{eq:dynamo2}). Omitting all the superscripts $^0$ and subscripts $_0$, and  bearing in mind that all the coefficients are evaluated at the reference state, we obtain the linearized dynamo equations from Eqs.~(\ref{eq:linear1}-\ref{eq:linear2}) when the $\alpha^2$-effects are neglected:
\begin{eqnarray}
\p{A'}{t} &=& (\alpha+c_0) B' \nonumber \\
 &&+ \beta\left( \frac{1}{r}\p{A'}{r} +\frac{\cot\theta}{r^2}\p{A'}{\theta} + \pp{A'}{r} + \frac{1}{r^2}\pp{A'}{\theta}\right), 
  \label{eq:l1} \\
\p{B'}{t} &=& a_1 A' + a_2 \p{A'}{r} + a_3 \p{A'}{\theta}  \nonumber\\
   &&  + c_1 B' + c_2 \p{B'}{r} + c_3 \p{B'}{\theta} + \beta\left(\pp{B'}{r} + \frac{1}{r^2} \pp{B'}{\theta}\right), \nonumber\\  
     \label{eq:l2}
\end{eqnarray}
where
\begin{eqnarray*}
 a_1 &=& \frac{\Omega}{r}(\cos\theta\hat{\Omega}_{r}-\hat{\Omega}_{\theta}\sin\theta) , \nonumber \\
 a_2 &=& -\Omega\hat{\Omega}_{\theta}\sin\theta , \nonumber \\
 a_3 &=& \frac{\Omega}{r}\hat{\Omega}_{r}\sin\theta, \\ 
 c_0 &=& \frac{2\beta_b}{B}\left( \frac{1}{r}\p{A}{r} +\frac{\cot\theta}{r^2}\p{A}{\theta} + \pp{A}{r} 
  + \frac{1}{r^2}\pp{A}{\theta}\right), \nonumber \\
 c_1 &=& \frac{2\beta_b}{r^2}(4\hat{B}_r+3\hat{B}_\theta\cot\theta)-\frac{1}{r^2}(2\beta_b+\beta)\csc^2\theta +\frac{\hat{\beta}}{r^2} \\
&& +\frac{2\beta_b}{r^2}(\hat{B}_r^2+\hat{B}_\theta^2) + \frac{2\beta_b}{B}\left(\pp{B}{r} + \frac{1}{r^2} \pp{B}{\theta}\right),  \nonumber \\
 c_2 &=& \frac{2\beta}{r}+ \frac{2\beta_b}{r}(1+2\hat{B}_r)+\frac{\hat{\beta}}{r} ,  \nonumber \\
 c_3 &=& \frac{2\beta}{r^2}\cot\theta+ \frac{2\beta_b}{r^2}(\cot\theta+2\hat{B}_\theta).
\end{eqnarray*}
We define $\hat{\Omega}_r$, $\hat{\Omega}_{\theta}$, and $\hat{\beta}$ is defined in Appendix A. Similarly, $\hat{B}_r\equiv\partial \ln B/\partial \ln r$, $\hat{B}_\theta\equiv\partial\ln B/\partial\theta$.

\subsection{Dispersion relation}

In order to investigate the linear instability, we try the wavelike solutions of the form
\begin{eqnarray}
  B' &=& B'_0\exp[i(\omega t-\vb{k}\cdot\vb{x})], \label{eq:solve1}\\
  A' &=& A'_0 \exp[i(\omega t-\vb{k}\cdot\vb{x})], \label{eq:solve2}
\end{eqnarray}
where $\omega$ is complex, while the other variables are real. 
Substituting Eqs.~(\ref{eq:solve1}-\ref{eq:solve2}) into Eqs.~(\ref{eq:l1}-\ref{eq:l2}), we obtain
\begin{eqnarray}
&& [i\omega+ \beta(ik_r/r +ik_\theta \cot\theta/r^2 +k^2)]A'_0 = (\alpha+c_0) B'_0,\nonumber\\ \label{eq:disp1}\\
&&  (i\omega-c_1 +ic_2k_r +ic_3 k_\theta + \beta k^2)B'_0 \nonumber \\
&& \hspace{9mm} = (a_1-ia_2 k_r -ia_3 k_\theta+ \alpha k^2) A'_0,  \label{eq:disp2}
\end{eqnarray}
where $k^2 \equiv k_r^2+k_\theta^2/r^2$. The typical length scale in the radial direction is $R_\sun$, while the typical angular scale in the latitude direction is $\pi$, therefore we have $k_r= 2m\pi /R_\sun $ and $k_\theta= 2\pi n/\pi = 2n$, where $m$ and $n$ are non-zero integers. For the fundamental wave, $m=n=1$. Consequently, 
$k=2[\pi^2 m^2+(R_\sun n/r)^2]^{1/2}/R_\sun$.

The dispersion equation can be obtained by multiplying Eq.~(\ref{eq:disp1}) and (\ref{eq:disp2}).
Defining
\begin{eqnarray*}
 \tilde{\omega} &=& i\omega +i\frac{1}{2}[(\beta/r+c_2)k_r+(\beta/r^2\cot\theta+c_3)k_\theta] \\
 && -\frac{1}{2}c_1+ \beta k^2, \\
 a_4 &=& a_2k_r+a_3k_\theta,\\
 c_4 &=& ({\beta}/{r}+c_2)k_r + (\beta\cot\theta/r^2+c_3)k_\theta, \\
 c_5 &=& \beta(k_r/r+k_\theta\cot\theta/r^2)(c_2k_r+c_3k_\theta), \\
 d_1 &=& c_1^2-c_4^2 +4c_5+4(\alpha+c_0) a_1, \\
 d_2 &=& 4c_1\beta(k_r/r+k_\theta\cot\theta/r^2) - 4(\alpha+c_0) a_4 -2c_1 c_4,
\end{eqnarray*}
we obtain
\begin{equation}
\tilde{\omega}^2 = \frac{1}{4}(d_1 + id_2).
\end{equation}
Its solutions are
\begin{equation}
\tilde{\omega}_{\pm} = \pm\frac{1}{2}\left( \sqrt{\frac{D+d_1}{2}}+i\sqrt{\frac{D-d_1}{2}}\right),
\end{equation}
where
\begin{equation}
 D=(d_1^2 + d_2^2)^{1/2}.
\end{equation}
The required solution is $\omega_+$ because we need a positive real part of $\omega$ to express the wave frequency:
\begin{equation}
  \omega =\omega_+= \omega_r + i\omega_i,
\end{equation}
where
\begin{eqnarray}
  \omega_r &=& \frac{1}{2}\left[ \sqrt{\frac{D-d_1}{2}} -(\beta/r+c_2)k_r-(\beta\cot\theta/r^2+c_3)k_\theta\right], \nonumber\\ \\
  \omega_i &=& -\frac{1}{2} \left( \sqrt{\frac{D+d_1}{2}} +c_1-2\beta k^2 \right).
\end{eqnarray}
The real part of $\omega$, $\omega_r>0$, determines the period of the dynamo wave, and the imaginary part of $\omega$, $\omega_i$, determines whether the dynamo wave grows ($\omega_i<0$) or decays ($\omega_i>0$).

The period $T$ is defined by
\[
   T=2\pi/\omega_r,
\]
and the growth rate $\Gamma$ is defined by
\[
  \Gamma = -\omega_i.
\]
These are two basic parameters of a dynamo. From their expressions we 
can see which mechanisms play a role.

\subsection{$\alpha$-$\Omega$ dynamo}\label{sect:b0}

When the buoyancy of the magnetic tube and the dynamic pressure 
of downflows are neglected, $\beta\equiv0$, the dispersion 
relation reduces to the usual version of the dynamo wave:
\begin{equation}
  \omega = \sqrt{\alpha}\left(\sqrt{\frac{D_0-a_1}{2}} - i \sqrt{\frac{D_0+a_1}{2}}\right),
    \label{eq:ao}
\end{equation}
where
\begin{equation}
  D_0 = [a_1^2+(a_2k_r+a_3k_\theta)^2]^{1/2}.
\end{equation}
Since $a_1$, $a_2$ and $a_3$ are related to $\Omega$, Eq.~(\ref{eq:ao}) shows that it is 
the $\alpha$- and $\Omega$-effects that determine the period and growth rate of the usual dynamo. 
That is why one calls it $\alpha$-$\Omega$ dynamo.

\begin{figure}
\plotone{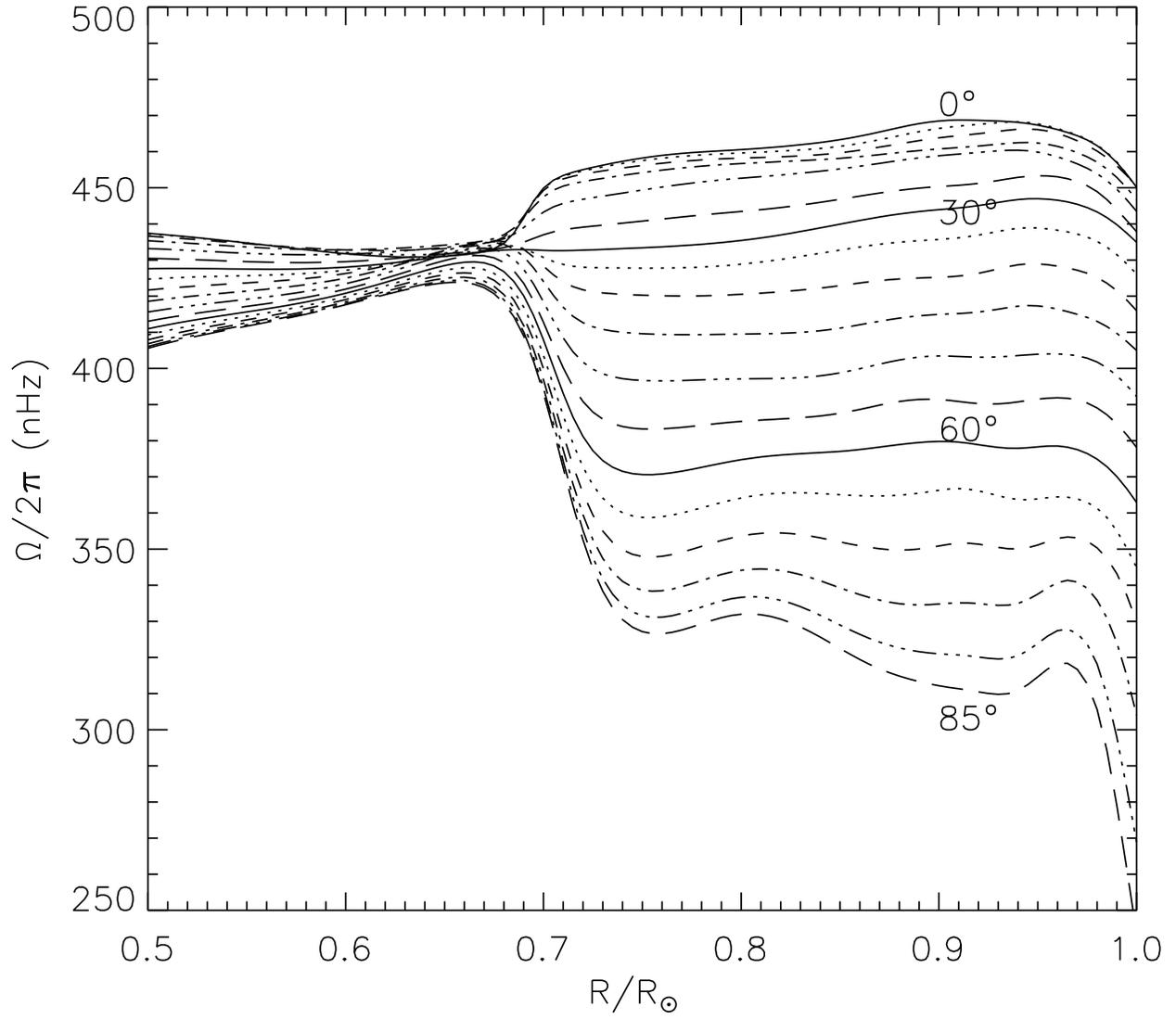}
\caption{
Solar rotation rate as a function of radius and latitude \citep{ABC98}.
\label{fig:rot}}
\end{figure}

When $\Omega$ is fixed by the observation, shown in Fig.~\ref{fig:rot}, the other adjustable 
parameters are wavenumbers $m$ and $n$, and the $\alpha$ parameter. Observations suggest that $m=1$
and $n=2$ for the equatorward branch, and $m=-1$ and $n=-2$ for the poleward branch, noticing 
that $k_r= 2\pi m/R_\sun$ and $k_\theta= 2\pi n/\pi$. The simplest $\alpha$ model is that $\alpha$ 
is a constant. The second simplest model is that $\alpha$ is some kind of function of radius and 
heliographic latitude. It is most likely that $\alpha$ depends upon dynamo magnetic fields.

\subsection{Anti-$\beta$ dynamo}

The parameter $\beta$ in the dynamo equations plays a role of dissipation. However, 
when $\beta$ is negative, the dissipation effect reverses, we have an anti-dissipation effect, 
called the anti-$\beta$ effect. When this effect dominates, the model is called an anti-$\beta$ dynamo.

In our model equations, we use $\beta$ to explicitly include the buoyancy and downflows effects 
on a magnetic tube. When $v_z$ is negative and the reference field $B$ is weaker than the critical 
field $B_c$ determined by $\beta_b+\beta_v=0$, $\beta$ is negative, we expect a positive growth rate for 
the tube field; when $v_z$ is negative and the reference field $B$ is stronger than the critical 
field $B_c$, $\beta$ is positive, we expect that the tube field damps; when $v_z$ is positive, 
the tube field should always damp; if we ignore the downflows, the tube should move upwards and 
the field at the specified place dissipates quickly, depending how strong the tube field is. 
We want to examine if we have correctly modeled these intuitive results.

The condition $\omega_r>0$ suggests
\begin{equation}
  (\beta/r+c_2)k_r+(\beta\cot\theta/r^2+c_3)k_\theta =0.
\end{equation}
If we demand $(\beta/r+c_2)k_r=0$ and $(\beta\cot\theta/r^2+c_3)k_\theta =0$, we find that the reference field satisfies
\begin{eqnarray}
  \hat{B}_{\theta}&=& -\frac{1}{2}\left(1+\frac{\beta}{\beta_b}\right)\cot\theta, \\
 \hat{B}_{r} &=& -\frac{1}{4}\left(2+3\frac{\beta}{\beta_b}\right)-\frac{\hat{\beta}}{4\beta_b}.
\end{eqnarray}
The amplitude is assumed to be proportional to the critical field $B_c$. We can further calculate the second-order derivatives 
which appear in the expression of $c_1$:
\begin{eqnarray*}
  \pp{B}{\theta} &=& B[\hat{B}_\theta^2+\frac{1}{2}(1+\beta/\beta_b)\csc^2\theta+(\beta/\beta_b)\hat{B}_\theta\cot\theta], \\
 \pp{B}{r} &=& \frac{B}{r^2}\left\{\hat{B}_r^2-\hat{B}_r + \frac{3}{2}\frac{\beta_v}{\beta_b}\hat{B}_r
  +\frac{1}{4}(\hat{\rho}-\hat{P})\hat{B}_r \right.\\
&&-\frac{1}{8}\frac{\beta_v}{\beta_b}(\hat{\rho}-\hat{P})(2+\hat{P})-\frac{3}{4}\frac{\beta_v}{\beta_b}(2+\hat{P})  \\
&& +\frac{1}{8}\left( 3\p{\hat{\rho}}{\ln r}-\p{\hat{P}}{\ln r}\right) 
 \left. -\frac{1}{8}\frac{\beta_v}{\beta_b}\left( \p{\hat{\rho}}{\ln r}-\p{\hat{P}}{\ln r}\right)\right\}, 
\end{eqnarray*}
where $\hat{\rho}=\partial\ln\rho/\partial\ln r=r/H_\rho$ and $\hat{P}=\partial\ln P/\partial\ln r=r/H_p$. 
Here $H_\rho$ and $H_p$ are density and pressure scale heights respectively.

We assume $A=frB$, where $f$ is a constant factor. A negative $f$ indicates that the poloidal field is 
in anti-phase with the toroidal field. Consequently, we obtain
\begin{eqnarray*}
  \p{A}{\theta} &=& fr\hat{B}_\theta, \\
  \p{A}{r} &=&f B(1+\hat{B}_r), \\
  \pp{A}{\theta} &=& fr \pp{B}{\theta}, \\
 \pp{A}{r} &=& \frac{2fB}{r}\hat{B}_r + fr \pp{B}{r}. 
\end{eqnarray*}
These expressions are used to calculate $c_0$.

Numerical calculations by using the standard solar model (Winnick et al. 2003) show that
\begin{eqnarray*}
 c_1 &<& 0, \\
 d_1 &\approx& c_1^2+4\alpha'a_1 \gg |d_2|, \\
 d_2 &\approx& 4 c_1\beta(k_r/r+k_\theta\cot\theta/r^2)+ 4\alpha'a_4,
\end{eqnarray*}
where $\alpha'=\alpha+c_0$. Since $c_0$ depends upon $B$, $\alpha'$ depends 
upon $B$ even if $\alpha$ does not depend upon $B$. The expression of $c_0$ 
given above provides us with a concrete nonlinear $\alpha$ model.
Consequently, the general dispersion relation reduces to
\begin{eqnarray}
  \omega &\approx& |-\beta(k_r/r+k_\theta\cot\theta/r^2)+\alpha'a_4/c_1|  \nonumber \\
   && +i (\beta k^2+ {\alpha'a_1}/{c_1}+d^2_2/16c_1^3).
\end{eqnarray}

Letting $\alpha'=0$, $\Gamma\approx-\beta k^2$, which confirms the qualitative requirements stated above.

In order to investigate the phase relationship between the poloidal and toroidal field, 
we set $\alpha=0$. Since $a_1<0$ near the surface, $c_1<0$ and the sign of $c_0$ is determined 
by the sign of $f$, we see that the term $\alpha'a_1/c_1>0$ when $f$ is positive and vice versa.
This shows that the growth rate $\Gamma\approx -(\beta k^2+ {\alpha'a_1}/{c_1})$ is larger when the two 
fields are in anti-phase with each other than when they are in phase. Obviously, this 
anti-phase relationship can be attributed to the nonlinear $\alpha$ effect. Physically, it 
is due to the nonlinear interaction between the poloidal and toroidal fields.

The observed equatorward and poleward branches of the solar magnetic field wave suggest
\begin{eqnarray}
   k_r &=& \left\{\begin{array}{ll}
     2\pi(+1)/R_\sun &  \mbox{ if $\theta> 45^\circ$},  \\[6pt]
     2\pi(-1)/R_\sun &  \mbox{ if $\theta< 45^\circ$},
    \end{array}\right.  \\
   k_\theta &=& \left\{\begin{array}{ll}
     2\pi(+2)/\pi &  \mbox{ if $\theta> 45^\circ$},  \\[6pt]
     2\pi(-2)/\pi &  \mbox{ if $\theta< 45^\circ$}.
    \end{array}\right.
\end{eqnarray}
The third term in the expression of the growth rate
\[
  \Gamma=-(\beta k^2+ {\alpha'a_1}/{c_1}+ d^2_2/16c_1^3)
\]
may explain the observed branching if we use the following $\alpha$ model:
\begin{equation}
   \alpha = \left\{\begin{array}{ll}
     +\alpha_0 &  \mbox{ if $\theta< 45^\circ$ and $k_r>0$ and $k_\theta>0$},  \\
     -\alpha_0 &  \mbox{ if $\theta< 45^\circ$ and $k_r<0$ and $k_\theta<0$},  \\
     -\alpha_1 &  \mbox{ if $\theta> 45^\circ$ and $k_r>0$ and $k_\theta>0$},  \\
     +\alpha_1 &  \mbox{ if $\theta> 45^\circ$ and $k_r<0$ and $k_\theta<0$},  
   \end{array}\right.  \label{eq:alfa}
\end{equation}
where $\alpha_0$ and $\alpha_1$ may slightly depend on $\theta$, but may strongly depend upon 
$r$ since the important $\theta$-dependence of $\alpha$ has been written down explicitly.
This is another nonlinear $\alpha$ effect, different from that specified by $c_0$. In fact, 
$\alpha$ may be represented by the perturbation magnetic field $\vb{B}'$ as follows 
(e.g., Seehafer et al. 2003):
\[
  \alpha =\frac{1}{3}\tau_{\s{cor}}\overline{\vb{B}'\cdot(\nabla\times\vb{B}')}.
\]
Using the wave-like solution, Eq.~(\ref{eq:wave}), one can obtain Eq.~(\ref{eq:alfa}).

We have two options to qualitatively explain why the poleward branch is 
weaker that the equatorward branch:
\begin{itemize}
  \item $\alpha_1<\alpha_0$,
  \item the observed poleward branch is the reflection wave, which is one part of the original wave.
\end{itemize}
Because of the phase delay of the poleward branch with respect to the equatorward branch, 
the second option is preferable.

%\subsection{Dynamo waves}

Solving Eq.~(\ref{eq:disp1}) for $A_0'$, we obtain
\[
  A' = R_1e^{i\Phi}B',
\]
where
\begin{eqnarray*}
  R_1 &\approx& |\alpha'|/[(d_2^2/16c_1^3)^2+\alpha'a_4/c_1)^2]^{1/2}, \\
  \Phi_1 &\approx& \arctan[\alpha' a_4/c_1,-(d_2^2/16c_1^3)].
\end{eqnarray*}
Using $R_1$ and $\Phi_1$, we can express the temporal evolution of the toroidal 
magnetic field $B$ and the poloidal vector potential $A$ in terms of the initial 
field $B_0$ for each magnetic flux tube as follows:
\begin{eqnarray*}
   B &=& B_0 e^{\Gamma t} \left\{\begin{array}{ll}
     \exp\left[i\left(\frac{2\pi t}{T}-\vb{k}\cdot\vb{x}+\Phi_0\right)\right] & \mbox{ if $\theta> 45^\circ$},  \\[6pt]
     \exp\left[i\left(\frac{2\pi t}{T}+\vb{k}\cdot\vb{x}+\Phi_0\right)\right] & \mbox{ if $\theta< 45^\circ$},
    \end{array}\right.  \\
   A &=& B_0R_1 e^{\Gamma t}e^{i\Phi_1} \left\{\begin{array}{ll}
     \exp\left[i\left(\frac{2\pi t}{T}-\vb{k}\cdot\vb{x}+\Phi_0\right)\right] & \mbox{ if $\theta> 45^\circ$} , \nonumber \\[6pt]
     \exp\left[i\left(\frac{2\pi t}{T}+\vb{k}\cdot\vb{x}+\Phi_0\right)\right] & \mbox{ if $\theta<45^\circ$},
    \end{array}\right. 
\end{eqnarray*}
where $\Phi_0$ is determined by the initial condition. Substituting these expressions into Eq.~(\ref{eq:b}), we obtain,
\begin{eqnarray}
   \vb{B} &=& B_0 e^{\Gamma t} \left\{ \begin{array}{rl} 
     \exp\left[i\left(\frac{2\pi t}{T}-\vb{k}\cdot\vb{x}+\Phi_0\right)\right] 
      [\vb{e}_\phi +\frac{R_1}{r}e^{i\Phi_1}\vb{a}^-] &  \\
               \mbox{ if $\theta> 45^\circ$}, & \\[6pt]
     \exp\left[i\left(\frac{2\pi t}{T}+\vb{k}\cdot\vb{x}+\Phi_0\right)\right]
      [\vb{e}_\phi +\frac{R_1}{r}e^{i\Phi_1}\vb{a}^+]  & \\
                \mbox{ if $\theta< 45^\circ$}, &
    \end{array} \right. \label{eq:wave}
\end{eqnarray}
where
\begin{eqnarray*}
   \vb{a}^- &=& (\cot\theta-ik_\theta)\vb{e}_r-(1-irk_r)\vb{e}_\theta, \\
   \vb{a}^+ &=& (\cot\theta+ik_\theta)\vb{e}_r-(1+irk_r)\vb{e}_\theta.
\end{eqnarray*}

For the equatorward branch, we can investigate how its peak propagates with time by setting
\begin{equation}
   2\pi t/T - k_r r -k_\theta \theta + \Phi_0 = (n-1)\pi,
\end{equation}
where $n$ equals to the integer part of $1+2t/T$. Since $k_r=\pi/R_\sun$ and $k_\theta=4$, 
we obtain the migration equation of the wave peak
\begin{equation}
  \theta = 2\pi/9 +[2t/T -(n-1)]\pi/4 ,
   \label{eq:eb}
\end{equation}
where we have assumed $\theta=40^\circ$ when $t=0$. Obviously, the integer $n$
plays a role of cycle number.

\begin{figure}
\plotone{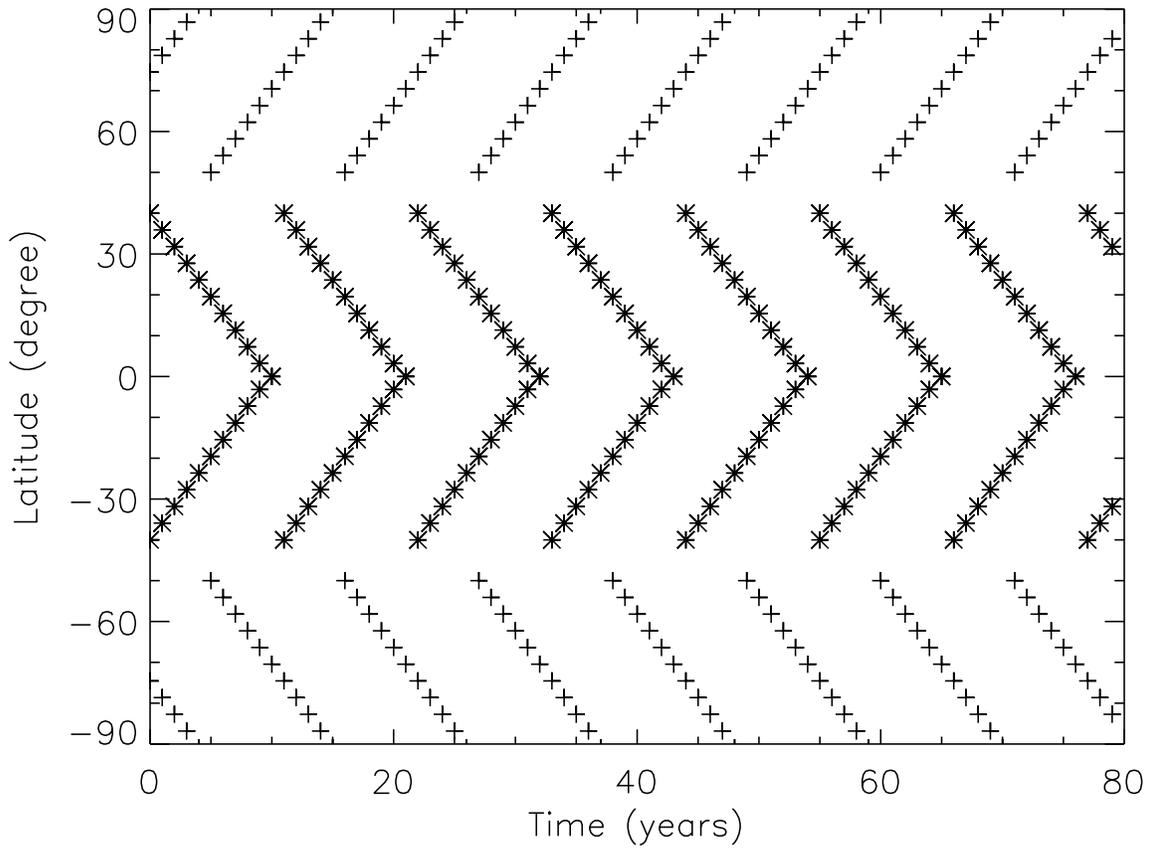}
\caption{
Heliographic latitude of the dynamo wave peak as a function of time\label{fig:butterfly}}
\end{figure}

For the poleward branch, we can only see the reflected wave because it propagates into the interior of the sun, $k_r=-2\pi/R_\sun$.
The reflection point is assumed to be located at the midpoint of the convection zone for the dynamo waves generated near the surface layers. This leads to $r_{\s{rfl}}\approx 0.86 R_\sun$. The phase velocity of the wave in the radial direction approximately equals to $100$ cm/s. Therefore, it takes about $t_0=6$ years for the wave to travel from the source region ($r_{\s{src}}=0.9995 R_\sun$) to the reflection point ($r_{\s{rfl}}=0.86 R_\sun$), and then to the surface ($r=R_\sun$). As a result, the wave peak propagation equation for the poleward branch reads as follows:
\begin{equation}
   2\pi (t+t_0)/T + k_r r + k_\theta \theta + \Phi_0 = (n'-1)\pi,
\end{equation}
where $n'$ equals to the integer part of $1+2(t+t_0)/T$. The solution of this equation is
\begin{equation}
  \theta = 5\pi/18 -[2(t+t_0)/T-(n'-1)]\pi/4,
   \label{eq:pb}
\end{equation}
where we have assumed $\theta=50^\circ$ when $t+t_0=0$. This implies a phase lag of $\pi/2$ with respect to the equatorward branch, as shown in Fig.~\ref{fig:butterfly}, which plots Eqs.~(\ref{eq:eb}) and (\ref{eq:pb}).

\section{Comparison with observations}\label{sect:s3}

In this section we try to compare the anti-$\beta$ dynamo model with relevant observations. The model parameters are listed as follows:
(i) the downflow velocity $v_z\approx -2.1\cdot 10^5$ cm/s is fixed by the 3d numerical simulations (Robinson et al 2003); (ii) we adjust the flux tube radius $a$ to obtain the observed cycle period $T\approx 22$ years; (iii) $\xi=1$, which means that the flux tube is totally located within the downflow regions; (iv) the reflection point of the dynamo wave is assumed to be located at the midpoint of the convection zone; (v) the reference magnetic fields vary from zero to the critical field $B_c$, which means that we should integrate all the observable quantities such as the cycle period over the reference field from zero to $B_c$; (vi) $\beta_t=\frac{1}{3}v_z^2 (\rho/P)^{1/2}2a$; (vii) $\alpha=0$; (viii) $f=-10^{-10}$.

\subsection{Period}

The $\beta$-effect dominates in the anti-$\beta$ dynamo model; the period $T$ near the surface can be approximately expressed as follows
\[
  T(a,\theta)=2\pi/\left\{ \begin{array}{ll}
    \int_0^{B_c} -\beta(2\pi/R_\sun+4\cot\theta/R_\sun^2)dB/B_c & \\
                 \hspace{4.4cm}\mbox{ if $\theta> 45^\circ$}, & \\
    \int_0^{B_c} \beta(2\pi/R_\sun+4\cot\theta/R_\sun^2)dB/B_c & \\
                 \hspace{4.4cm} \mbox{ if $\theta< 45^\circ$}. &
    \end{array} \right.
\]
If one fixes $T\approx 22$ years, one finds a=(1.15, 1.1, 1.05, ...,0.65, 0.59, 0.53, 0.45, 0.37, 0.27, 0.15) Mm for
$\theta$=(5$^\circ$, 10$^\circ$, 15$^\circ$,..., 55$^\circ$, ..., 85$^\circ$), respectively.
This is comparable to 2 Mm, the typical size of flux tubes observed in the low latitude region.

\subsection{Growth rate}

Similarly, the growth rate $\Gamma$ of a flux tube near the surface can be approximately expressed as follows
\[
  \Gamma(a)= \int_0^{B_c} -\beta k^2dB/B_c \mbox{ for $0^\circ<\theta < 90^\circ$},
\]
where $k^2=(2\pi+4)^2/R_\sun^2$.
For the parameters given above, one finds 
$\Gamma^{-1}$=(0.46, 0.48, 0.51, 0.53, 0.56, 0.59, 0.62, 0.66, 0.71, 0.76, 0.82, 0.94, 1, 1.18, 1.44, 1.97, 3.54) 
years. This shows that the growth time of a typical flux tube of size $a=2$ Mm is about several months.

The flux tubes should stay in the convection zone long enough (i.e., for a time larger than or equal to their growth time) 
for them to be amplified by the dynamo action. Since our formulation has already included the buoyant dissipation 
effect, a positive growth rate measured by the growth time indicates that the downflows can inhibite the buoyant 
rise of the flux tube on the time scale specified by the growth time. Here we assume that the flux tube is so distorted
that it is totally immersed in the downflow regions by setting $\xi=1$. From Fig.~5 of Fan et al (2003), we can see 
that only a small fraction of their flux tube is immersed in the downflow regions, which means that their $\xi$ parameter is much smaller than the value we use. Therefore, our results are not in conflict with their 3D simulation, which shows that the pumping has almost no effect in inhibiting the buoyant rise of the flux tube when the energy of the tube is larger than that of the surrounding environment. If the energy of the tube is comparable to that of the surrounding, it still rises to the surface, but being largely distorted. If we use a small $\xi$ parameter value, we can obtain similar results.

\subsection{Active regions}

Here we have two reasons to explain that the strong activity may be confined to low heliographic latitudes $|\Phi|\le 35^\circ$ of the equatorward branch, as observed:
\begin{description}
\item{(i)} growth rate at low latitude is larger than that at high latitude;
\item{(ii)} what we observe at high heliographic latitudes are the reflected waves. They must be weaker than the original waves since the reflection is not necessarily complete.
\end{description}

\subsection{Phase dilemmas}

\begin{description}
  \item{(1)} Eq.~(\ref{eq:wave}) shows that the phase difference between the poloidal and toroidal field is $\Phi_1$. Calculations show $\Phi_1\approx 170^\circ$ for $0^\circ<\theta<180^\circ$. This is in good agreement with the observation: the poloidal field is almost in antiphase with the toroidal component;
 \item{(2)} The $\pi/2$ phase lag of the poleward branch with respect to the equatorward branch can be explained by the time lag 
of 6 years for the former to travel from the source region to the reflection point, and then from the reflection point to the surface.
 \end{description}

\subsection{Butterfly diagram}

Fig.~\ref{fig:butterfly} depicts the latitude of the peak of the dynamo wave as a function of time. Only when we select the mid-point of the convection zone as the reflection point, the time delay of the poleward branch is about 6 years. This is a bit arbitrary.

\subsection{Cyclic variations of solar global parameters and helio-seismic frequencies}

We construct models of the structure and evolution of the Sun which include variable 
magnetic fields and turbulence. The magnetic effects are: (1) magnetic pressure, (2) magnetic energy, and  
(3) magnetic modulation to turbulence. The  effects of turbulence are: 
(1) turbulent pressure, (2) turbulent kinetic energy, and (3) turbulent 
inhibition of the radiative energy loss of a convective eddy, and 
(4) turbulent generation of magnetic fields. Using these ingredients we 
construct five types of solar variability models (including the standard solar model) 
with magnetic effects. These models are in part based on three-dimensional numerical 
simulations of the superadiabatic layers near the surface of the Sun.
The models are tested with several sets of observational data, namely, the changes of 
(1) the total solar irradiance, (2) the photospheric temperature, (3) radius, (4) the position 
of the convection zone base,  (5) low- and medium-degree solar oscillation frequencies.
We find that turbulence plays a major role in solar variability and only a
 model which includes a magnetically-modulated 
turbulent mechanism can agree with all the current available observational data.
We find that because of the somewhat poor quality of all observations (other than the
helioseismological ones), we need all data sets in order to restrict the range of
models. These studies, presented in Li et al (2003), strongly suggest a near-surface dynamo model.
The anti-$\beta$ dynamo, which operates near the surface, may meet this need.

\section{Conclusions}\label{sect:s4}

Helioseimic observations of the cyclic variations of low- and medium-degree solar 
oscillation frequencies are very precise. The cyclic variation of the position 
of the convection zone base, obtained by helioseismic inversions (Basu \&\ Antia 2000; 
Basu 2002), is also very accurate. 
These high-precision observations seem to exclude the location of
strong magnetic fields at the base of the convection zone or in the overshoot
layer, as is generally assumed on the basis of stability against magnetic
buoyancy and equipartition arguments. This is one of the main motivations to
reconsider dynamos located in the convection zone. The second motivation comes
from trying to reproduce the observed cyclic variations of the total solar
irradiance \cite{FL98}, the solar effective temperature \cite{GL97}, and the
solar radius (See \S2.3 of Li et al 2003 for a brief review). Extensive numerical 
experiments \cite{LS01} show that only magnetic fields located in the convection 
zone can simultaneously produce the observed cyclic variations of these three global 
solar parameters. The third motivation originates from numerical simulations of the solar
convection zone \cite{CS89,N99,RDLSKCG03}, because they indicate a major presence of
downward-moving plumes of high velocity.

However, convection zone dynamos have to face the magnetic buoyancy instability
problem. The best way to solve this problem is to incorporate the buoyancy into the dynamo
equations. This forces us to treat the mean field in a flux tube. To include the
dynamic pressure of a flow, the flux tube should be regarded as a body. As
a field, both the cyclonic convection ($\alpha$-effect) and differential
rotation ($\Omega$-effect) play a role. As a body, the tube experiences not only
a buoyant force, but also a dynamic pressure of downflows above the tube. We
show that these two dynamic effects can be incorporated into the dynamo
equations by converting them in two diffusion terms.

We analyzed and solved the extended dynamo equations in the linear approximation
by adopting the observed solar internal rotation rate, and by assuming a downflow
suggested by numerical simulations of the solar convection zone. The results are
in agreement with the basic properties of the solar cycle, and give a theoretical
basis to understand the variations of solar irradiance, temperature, radius
and helioseismic frequencies in the course of the magnetic cycle.

Finally, the model results give a theoretical basis to interpret the observed
cyclic variations of the total solar irradiance, the solar effective
temperature, the solar radius, and the helioseismic frequencies (Li et 
al. 2003; Le et al 2002; Li \&\ Sofia, 2001; Sofia \&\ Li, 2001).

\acknowledgements
We want to thank Dr. Sarbani Basu for her kindly providing us with rotation 
rate data, Dr. Frank J. Robinson for his help, and Prof. Pierre Demarque for 
useful discussions. This work was supported  by  NASA  grant 899-10633 and NSF grants 
ATM 9303023 and ATM 0206130.

\appendix

\section{A. Dynamo equations in spherical coordinates}\label{app:a}

In this paper we assume azimuthal symmtry, which means that all quantities do not depend upon the azimuthal coordinate $\phi$.

Using Eq.~(\ref{eq:b}) to expand Eq.~(\ref{eq:model}), we obtain
\begin{equation}
  \p{B}{t}\vb{e}_{\phi} + \nabla\times \left[\left(\p{A}{t}-\alpha B-\beta\nabla^2 A\right) \vb{e}_{\phi}\right]
    = \nabla\times\{U\vb{e}_{\phi}\times [\nabla\times (A\vb{e}_{\phi})]
    +\alpha \nabla\times (A\vb{e}_{\phi})
    -\beta\nabla\times (B \vb{e}_{\phi})\}, \label{eq:ba}
\end{equation}
where we have used 
\begin{equation}
   \vb{U} = \vb{\Omega}\times \vb{r} = r\Omega \sin\theta \vb{e}_{\phi}=U\vb{e}_{\phi}.
\end{equation}
We have also used the mathematical equality
\begin{eqnarray}
   \nabla\times\nabla\times (A\vb{e}_{\phi}) &=& \nabla(\nabla\cdot A\vb{e}_{\phi}) - \nabla^2 (A\vb{e}_{\phi}) \nonumber\\
    &=& - \vb{e}_{\phi} \nabla^2 A.
\end{eqnarray}
The second equality hold well when $A$ does not depend upon $\phi$.

We use the following mathematical equality to calculate the curl of a vector field $\vb{F}$,
\begin{equation}
  \nabla\times \vb{F} = \frac{1}{r\sin\theta}\left[\p{}{\theta}(\sin\theta F_{\phi})-\p{F_{\theta}}{\phi} \right]\vb{e}_r
    +\frac{1}{r}\left[\frac{1}{\sin\theta}\p{F_r}{\phi}-\p{}{r}(rF_\phi) \right] \vb{e}_{\theta}
    +\frac{1}{r}\left[\p{}{r}(rF_{\theta})- \p{F_r}{\theta}\right]\vb{e}_{\phi}.
\end{equation}
Since we assume that the system is in azimuthal symmetry, we know
\begin{equation}
  \nabla\times(F\vb{e}_{\phi}) = \frac{1}{r\sin\theta}\p{}{\theta}(\sin\theta F) \vb{e}_r -\frac{1}{r}\p{}{r}(r F) \vb{e}_{\theta}
    \label{eq:ephi}
\end{equation}
for $\vb{F}=F(r,\theta) \vb{e}_{\phi}$. Using these formula in Eq.~(\ref{eq:ba}), we obtain
\begin{equation}
 \left(\p{B}{t}-\frac{M}{r}\right)\vb{e}_{\phi} + \nabla\times \left[\left(\p{A}{t}-\alpha B-\beta\nabla^2A\right) \vb{e}_{\phi}\right]  
    =0, \label{eq:ba1}
\end{equation}
where
\begin{eqnarray}
 M &=& r\nabla\times\left\{\left[\frac{U}{r}\p{}{r}(rA)+\frac{\alpha}{r\sin\theta}\p{}{\theta}(\sin\theta A)-\frac{\beta}{r\sin\theta}\p{}{\theta}(\sin\theta B)   \right]\vb{e}_r \right. \nonumber \\
 && \left. + \left[\frac{U}{r\sin\theta}\p{}{\theta}(\sin\theta A) -\frac{\alpha}{r}\p{}{r}(r A)+\frac{\beta}{r}\p{}{r}(r B) \right]\vb{e}_{\theta}\right\}\cdot\vb{e}_\phi  \nonumber \\
 &=& \p{}{r}\left[\frac{U}{\sin\theta}\p{}{\theta}(\sin\theta A) - \alpha\p{}{r}(rA)
  +\beta\p{}{r}(rB)\right] \nonumber \\
 && -\p{}{\theta}\left[\frac{U}{r}\p{}{r}(rA) +\frac{\alpha}{r\sin\theta}\p{}{\theta}(\sin\theta A)
 -\frac{\beta}{r\sin\theta}\p{}{\theta}(\sin\theta B) \right] \nonumber\\
 &=& \p{}{r}\left[U\p{A}{\theta} +(U\cot\theta -\alpha) A -\alpha r\p{A}{r} +\beta B  +\beta r\p{B}{r}\right] \nonumber \\
 && -\p{}{\theta}\left[U\p{A}{r} +\frac{\alpha}{r}\p{A}{\theta}+\frac{U+\alpha\cot\theta}{r} A
 -\frac{\beta}{r}\p{B}{\theta}-\frac{\beta\cot\theta}{r}B \right] \nonumber\\
 &=& \left\{\p{}{r}(U\cot\theta-\alpha)-\frac{1}{r}\p{}{\theta}[U+\alpha\cot\theta)]\right\} A   
    + \left\{(U\cot\theta-\alpha) -\p{}{r}(\alpha r)-\p{U}{\theta}  \right\} \p{A}{r} \nonumber \\
 &&   - \left\{\frac{U+\alpha\cot\theta}{r}+ \p{}{\theta}\left(\frac{\alpha}{r}\right)-\p{U}{r} \right\} 
   \p{A}{\theta} -\alpha r\pp{A}{r} - \frac{\alpha}{r}\pp{A}{\theta} \nonumber \\
&& +\left[\p{\beta}{r}+\p{}{\theta}\left(\frac{\beta\cot\theta}{r}\right)\right]B 
  +\left[\beta+\p{}{r}(\beta r)\right]\p{B}{r} +\left[\frac{\beta\cot\theta}{r} +\p{}{\theta}(\beta/r)\right]\p{B}{\theta} \nonumber\\
&&  +\beta r\pp{B}{r} +\frac{\beta}{r}\pp{B}{\theta}.
\end{eqnarray}
Eq.~(\ref{eq:ephi}) shows that the second term of Eq.~(\ref{eq:ba1}) on the left hand side is perpendicular to the first term. Consequently, Eq.~(\ref{eq:ba1}) is equivalent to the following two scalar equations:
\begin{eqnarray}
 \p{A}{t} &=& \alpha B + \beta\left( \frac{1}{r}\p{A}{r} + \frac{\cot\theta}{r^2}\p{A}{\theta} + \pp{A}{r} + \frac{1}{r^2}\pp{A}{\theta}\right), \\
 \p{B}{t} &=& a_1 A + a_2 \p{A}{r} + a_3 \p{A}{\theta} -\alpha \left(\pp{A}{r}+\frac{1}{r^2} \pp{A}{\theta}\right) + b_1 B + b_2 \p{B}{r} 
   + b_3\p{B}{\theta} \nonumber \\
&&   + \beta \left(\pp{B}{r} + \frac{1}{r^2} \pp{B}{\theta}\right),
\end{eqnarray}
where we have used
\begin{eqnarray}
   \nabla^2 A &=& \frac{1}{r^2}\p{}{r}\left(r^2\p{A}{r}\right)+\frac{1}{r^2\sin\theta}\p{}{\theta}\left(\sin\theta\p{A}{\theta}\right) \nonumber \\
 &=& \frac{2}{r}\p{A}{r} +\frac{\cot\theta}{r^2}\p{A}{\theta} + \pp{A}{r} + \frac{1}{r^2}\pp{A}{\theta}
\end{eqnarray}
and defined
\begin{eqnarray}
 a_1 &=& \frac{1}{r} \left[\p{}{r}(U\cot\theta-\alpha)-\frac{1}{r}\p{}{\theta}(U+\alpha\cot\theta)\right] \nonumber\\
   &=&\frac{1}{r}\left(\Omega\cos\theta\hat{\Omega}_{r}-\Omega\hat{\Omega}_{\theta}\sin\theta+\frac{\alpha}{r}\csc^2\theta 
    -\p{\alpha}{r}     -\frac{1}{r}\cot\theta\p{\alpha}{\theta}\right) , \nonumber \\
 a_2 &=& \frac{1}{r} \left\{(U\cot\theta-\alpha) -\p{}{r}(\alpha r)-\p{U}{\theta}  \right\}
   = -\Omega\hat{\Omega}_{\theta}\sin\theta-\frac{2\alpha}{r}-\p{\alpha}{r} , \nonumber \\
 a_3 &=& - \frac{1}{r}\left\{\frac{U+\alpha\cot\theta}{r}+ \p{}{\theta}\left(\frac{\alpha}{r}\right)-\p{U}{r} \right\}
   =\frac{1}{r}\left(\Omega\hat{\Omega}_{r}\sin\theta -\frac{\alpha\cot\theta}{r}-\frac{1}{r}\p{\alpha}{\theta}\right), \\ 
 b_1 &=& \frac{1}{r} \left[\p{\beta}{r}+\p{}{\theta}\left(\frac{\beta\cot\theta}{r}\right)\right]
   =\frac{2\beta_b}{rB}\left(\p{B}{r}+ \frac{\cot\theta}{r}\p{B}{\theta}\right)-\frac{\beta\csc^2\theta}{r^2}
    +\frac{\hat{\beta}}{r^2}, \nonumber \\
 b_2 &=& \frac{1}{r}\left[\beta+\p{}{r}(\beta r)\right]=\frac{2\beta}{r}+ \frac{2\beta_b}{B}\p{B}{r}
   +\frac{\hat{\beta}}{r}, \nonumber \\
 b_3 &=& \frac{1}{r}\left[\frac{\beta\cot\theta}{r} +\p{}{\theta}(\beta/r)\right]=\frac{1}{r^2}\left(\beta\cot\theta
  +\frac{2\beta_b}{B}\p{B}{\theta}\right). \nonumber 
\end{eqnarray}
The valid condition for the second equality sign is $B\ll \sqrt{8\pi P}$. We define $\hat{\Omega}_r=\partial\ln\Omega/\partial\ln r$ and
$\hat{\Omega}_{\theta}=\partial\ln\Omega/\partial\theta$. We also define
\[
  \hat{\beta}\equiv \left(\p{\beta}{\ln r}\right)_B =\frac{\beta_t}{2}\left(\p{\ln\rho}{\ln r}-\p{\ln P}{\ln r}\right)-\beta_b\left(2+\frac{3}{2}\p{\ln P}{\ln r}-\frac{1}{2}\p{\ln\rho}{\ln r}\right)
   +\frac{\beta_v}{2}\left(\p{\ln P}{\ln r}-\p{\ln\rho}{\ln r}\right).
\]

\section{B. Linearization of Dynamo equations}\label{app:b}

In order to linearize Eqs.~(\ref{eq:dynamo1}-\ref{eq:dynamo2}), we decompose $B$ and $A$ into two parts:
\begin{eqnarray}
  B &=& B_0 + B', \label{eq:b1} \\
  A &=& A_0 +A', \label{eq:a1}
\end{eqnarray}
where $B_0$ and $A_0$ represent the reference state of the system, while $B'$ and $A'$ represents the perturbations of $B$ and $A$ near $B_0$ and $A_0$. Substituting Eqs.~(\ref{eq:b1}-\ref{eq:a1}) into Eqs.~(\ref{eq:dynamo1}-\ref{eq:dynamo2}), we obtain two linear equations that govern $B'$ and $A'$,
\begin{eqnarray}
  \p{A'}{t} &=& (\alpha+c_0) B' + \beta_0\left( \frac{1}{r}\p{A'}{r} +\frac{\cot\theta}{r^2}\p{A'}{\theta} + \pp{A'}{r} + \frac{1}{r^2}\pp{A'}{\theta}\right), \label{eq:linear1}\\
  \p{B'}{t} &=& a_1^0 A' + a_2^0 \p{A'}{r} + a_3^0 \p{A'}{\theta} -\alpha\left(\pp{A'}{r}+\frac{1}{r^2} \pp{A'}{\theta}\right) \nonumber\\
   &&  + c_1 B' + c_2 \p{B'}{r} + c_3 \p{B'}{\theta} + \beta_0\left(\pp{B'}{r} + \frac{1}{r^2} \pp{B'}{\theta}\right),  \label{eq:linear2}
\end{eqnarray}
where the superscript $^0$ or the subscript $_0$ means that the quantities are evaluated at the reference state, for example, $a_1^0 = a_1|_{B=B_0, A=A_0}$, and
\begin{eqnarray}
 c_0 &=& \frac{2\beta^0_b}{B_0}\left( \frac{1}{r}\p{A_0}{r} +\frac{\cot\theta}{r^2}\p{A_0}{\theta} + \pp{A_0}{r} + \frac{1}{r^2}\pp{A_0}{\theta}\right), \nonumber \\
 c_1 &=& b_1^0 +  \p{b_1^0}{B_0} B_0  + \p{b_2^0}{B_0} \p{B_0}{r} + \p{b_3^0}{B_0} \p{B_0}{\theta} + \frac{2\beta^0_b}{B_0}\left(\pp{B_0}{r} + \frac{1}{r^2} \pp{B_0}{\theta}\right),  \nonumber \\
 c_2 &=& b_2^0+ \p{b_1^0}{B'_0} B_0  + \p{b_2^0}{B'_0} \p{B_0}{r} + \p{b_3^0}{B'_0} \p{B_0}{\theta},  \nonumber \\
 c_3 &=& b_3^0 + \p{b_1^0}{B''_0} B_0  + \p{b_2^0}{B''_0} \p{B_0}{r} + \p{b_3^0}{B''_0} \p{B_0}{\theta}. \nonumber
\end{eqnarray}
Here $B'_0=\partial B_0/\partial r$ and $B''_0 =\partial B_0/\partial \theta$. Using the expressions for b's given in Appendix~\ref{app:a},
we can calculate their derivatives with respect to $B_0$, $B_0'$, and $B_0''$: 
\begin{eqnarray}
 \p{b_1^0}{B_0} &=& \frac{2\beta^0_b}{rB^2_0}\left(\p{B_0}{r}+ \frac{\cot\theta}{r}\p{B_0}{\theta}\right)
  -\frac{2\beta^0_b}{B_0}\frac{\csc^2\theta}{r^2}
 -\frac{2\beta_b}{r^2B_0}\left(2+\frac{3}{2}\p{\ln P}{\ln r}-\frac{1}{2}\p{\ln\rho}{\ln r}\right), \nonumber \\
 \p{b_2^0}{B_0} &=& \frac{4\beta^0_b}{rB_0} + \frac{2\beta^0_b}{B^2_0}\p{B_0}{r}-\frac{2\beta_b}{rB_0}\left(2+\frac{3}{2}\p{\ln P}{\ln r}-\frac{1}{2}\p{\ln\rho}{\ln r}\right),   \nonumber \\
 \p{b_3^0}{B_0} &=& \frac{1}{r^2}\left(\frac{2\beta^0_b}{B_0}\cot\theta+\frac{2\beta^0_b}{B^2_0}\p{B_0}{\theta}\right), \nonumber \\
 \p{b_1^0}{B'_0} &=& \frac{2\beta^0_b}{rB_0}, \nonumber \\
 \p{b_2^0}{B'_0} &=&  \frac{2\beta^0_b}{B_0}, \nonumber \\
 \p{b_3^0}{B'_0} &=& 0, \nonumber \\
 \p{b_1^0}{B''_0} &=& \frac{2\beta^0_b}{rB_0}\frac{\cot\theta}{r}, \nonumber \\
 \p{b_2^0}{B''_0} &=& 0, \nonumber \\
 \p{b_3^0}{B''_0} &=& \frac{2\beta^0_b}{r^2B_0}. \nonumber
\end{eqnarray}

\end{document}